# On smooth or 0/1 designs of the fixed-mesh element-based topology optimization

Xiaodong Huang

Faculty of Science, Engineering and Technology, Swinburne University of Technology, Hawthorn, VIC 3122, Australia

[*]E-mail: xhuang@swin.edu.au

Tel: +61-3-9214 5633

**Abstract**

The traditional element-based topology optimization based on material penalization typically aims at a 0/1 design. Our numerical experiments reveal that the compliance of a smooth design is overestimated when material properties of boundary intermediate elements under the fixed-mesh finite element analysis are interpolated with a material penalization model. This paper proposes a floating projection topology optimization (FPTO) method for seeking a smooth design using the ersatz material model or a 0/1 design using a material penalization model. The proposed floating projection constraint combining with the upper and lower bounds heuristically simulates 0/1 constraints of design variables in the original discrete optimization problem. Numerical examples demonstrate the capability of the proposed element-based topology optimization approach in obtaining 0/1 or smooth designs for 2D and 3D compliance minimization problems. The proposed topology optimization approach can be easily implemented under the framework of the fixed-mesh finite element analysis and provides an alternative way to form explicit topologies of structures, especially when the ersatz material model is adopted.

# 1. Introduction

Topology optimization seeks the best material distribution within the design domain so that the resulting structure achieves the best or prescribed performance. Various topology optimization methods have been developed since the seminal paper on the homogenization method by Bensøe and Kikuchi [1] in 1988. The most popular gradient topology optimization includes the boundary-based approaches, such as the level-set (LS) method [2-5], and the element-based approaches, such as the solid isotropic material penalization (SIMP) method [6-8] and the bi-directional evolutionary structural optimization (BESO) method [9, 10].

The level-set method is one of the representative boundary-based topology optimization methods by capturing the change of structural topology by merging holes [2-5]. The design variables are defined based on the boundary, which is represented by the zero level contour of the level-set function. Updating the level-set function via the solution of Hamilton-Jacobi equation drives the variation of a structural boundary as well as topology. The other boundary-based topology optimization methods include the moving morphable components (MMC) method [11, 12] and the feature-driven optimization method [13]. However, the optimized design resulting from the boundary-based topology optimization algorithms is highly dependent on the initial guess design (e.g., number and location of holes), and the adopted regularization technique [14]. The LS method and its variants have significantly gained in popularity due to their promise to operate on the clearly defined boundaries throughout the optimization process [15-20]. In most cases, the structural topology is still mapped to the fixed finite element mesh to avoid re-meshing during the optimization process. As a result, the optimized design under the fixed mesh contains one layer of intermediate boundary elements whose material properties are interpolated with the ersatz material model.

The density method is a popular element-based topology optimization algorithm. Instead of using the microstructure of a material in the homogenization method [1], the density method [6-8] assumes that each element is composed of isotropic porous materials featured with their densities, $\rho$, (1 for solid and 0 for void). The relationship between the material property and the elemental density can be represented by the material penalization scheme, such as a power-law function or others [6, 21, 22]. Physically, materials with the density between 0 and 1 are unfavorable in the sense that the stiffness obtained is small compared to the cost of the material. The optimal solution theoretically tends to a solid/void design, but in practice, the elements with intermediate density ($0 < \rho < 1$) in final designs seem to be unavoidable. The advanced numerical algorithms including the continuation method [23, 24], the projection technique [25-28] have been developed for avoiding a local optimum with "grey" areas. Recently, Liang and Chen [29] proposed a new topology optimization method via sequential integer programming and Canonical relaxation algorithm to achieve 0/1 solutions by combining with the move limit strategy.

Based on the existence of a 0/1 solution ensured by material penalization, the gradient-based BESO method [9, 10] heuristically updates the densities of elements with discrete values, 0 and 1. Different from the SIMP method, BESO normally starts from full design and gradually decreases the total volume of the structure until the objective volume is achieved. Although BESO has its advantage in providing a pure 0/1 design for various topology optimization problems [30, 31], the zig-zag boundary has to be post-processed. Recently, Da et al. [32] developed a heuristic evolutionary topology optimization for obtaining a smooth design based on the SIMP model.

The element-based topology optimization method has its advantage in freely digging holes inside the design domain. In the development of the element-based topology optimization method, seeking a pure 0/1 design becomes its priority [25-28]. Nevertheless, such a pure 0/1 design leads to zig-zag boundaries which is less practical for fabrication. Therefore, the element-based topology optimization method was often suggested for the concept design of structures followed by the shape optimization for achieving smooth boundaries. The prominent question is whether the element-based topology optimization can provide a smooth design, which is similar to that of the boundary-based method. It is noticed that, when topology optimization conducts under the fixed-mesh finite element analysis, the boundary-based method simulates intermediate boundary elements using the ersatz material model, but the element-based method uses the material penalization model. Therefore, it is necessary to explore the relationship between the material interpolation scheme and a smooth or 0/1 design in the next section.

## 2. Smooth or 0/1 design vs. material interpolation scheme

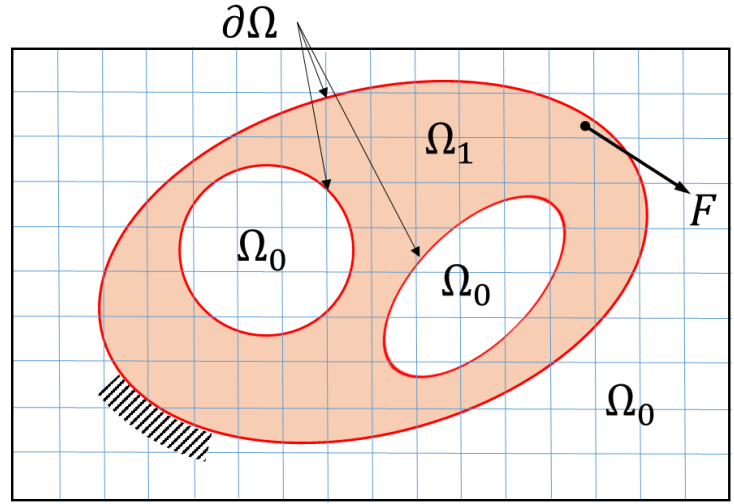

Fig. 1 Illustration on topology optimization problems based on the fixed-mesh finite element analysis.

A typical topology optimization problem as shown in Fig. 1 can be expressed by

$$\min_{\vartheta}: f(\vartheta)$$
$$\text{s.t.}: \int_{\Omega} \vartheta dV - V_0 \leq 0 \qquad (1)$$
$$\begin{cases} \vartheta_j = 1 \text{ when } \vartheta_j \in \Omega_1 \\ \vartheta_j = 0 \text{ when } \vartheta_0 \in \Omega_0 \end{cases}$$

where $f(\vartheta)$ is the objective function and $\vartheta = \{\vartheta_1, \vartheta_2, ...\}$. $\Omega = \Omega_1 \cup \Omega_0$ denotes the design domain in $\mathbf{R}^2$ or $\mathbf{R}^3$, which contains an infinite number of pixels, $\vartheta_j$. $\Omega_1$ and $\Omega_0$ denote the solid and void domains, respectively. $V_0$ is the constraint value of the volume.

To achieve a smooth design within the framework of the fixed-mesh finite element analysis illustrated in Fig. 1, the topology optimization problem, e.g., minimizing compliance, can be rewritten as

$$\min_{x}: C(x) = \mathbf{f}^T \mathbf{u}$$
$$\text{s.t.}: \sum x_i V_i - V_0 \leq 0 \qquad (2)$$
$$\begin{cases} x_i = 1 \text{ when } x_i \in \Omega_1 \\ x_i = x_{min} \text{ when } x_i \in \Omega_0 \\ x_{min} < x_i < 1 \text{ when } x_i \in \partial\Omega \end{cases}$$

where $x = \{x_1, x_2, ..., x_i, ..., x_N\}$ are the design variables associated with finite elements. $x_{min}$ denotes void with a small positive value, e.g., $10^{-3}$, to avoid the singularity of the problem. $\partial\Omega$ denotes the boundary between the solid and void domains. The original pixel-wise design problem with infinite design variables converts to a piecewise design problem with finite design variables. The final design of the above topology optimization allows one layer of intermediate elements at the boundary domain, $\partial\Omega$, and solid or void elements for the remaining domains.

Instead, a topology optimization problem is often formulated to seek a pure 0/1 design as

$$\min_{x}: C(x) = \mathbf{f}^T\mathbf{u}$$
$$\text{s.t.}: \sum x_i V_i - V_0 \leq 0 \qquad (3)$$
$$x_i \in [x_{min}, 1]$$

where only solid or void elements are allowed for the whole design domain.

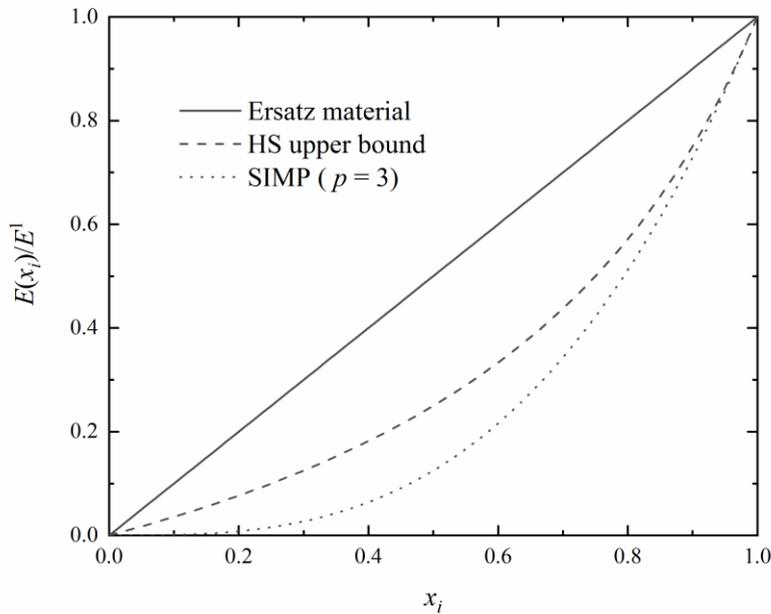

Fig. 2 The relationship between the material property and design variables under various interpolation schemes.

To solve the topology optimization problems expressed by eqns. (2) and (3) mathematically, the design variables can be relaxed to attain any values between $x_{min}$ and 1 (e.g., $x_{min} \leq x_i \leq 1$). As a result, topology optimization problems for a smooth or 0/1 design become the same one except for the interpretation of intermediate boundary elements as explained below. The material property, $E(x_i)$, is artificially interpolated for any values of design variables, $x_i$ between $x_{min}$ and 1. The material interpolation schemes include the linear interpolation scheme, i.e., the ersatz material model [1, 2, 4], and the material penalization schemes, including the Hashin-Shtrikman (HS) upper bound model [33, 34] and the well-known SIMP model [6, 21].

$$E(x_i) = \begin{cases} x_i E^1 & \text{for the ersatz material} \\ \dfrac{x_i}{3 - 2x_i} E^1 & \text{for HS upper bound (in 2D)} \\ x_i^p E^1 & \text{for the SIMP model} \end{cases} \quad (4)$$

where $E^1$ denotes Young's modulus of the solid. $p > 1$ is the penalty exponent for the SIMP model and $p = 3$ is used throughout this paper. Fig. 2 depicts the relationships between Young's modulus and design variables for those material interpolation schemes.

To identify which material model is suitable for a smooth design, we conduct the numerical analysis for a simple two-bar structure shown in Fig. 3(a). When the structure is discretized with an irregular mesh (the typical element size is approximately 1), the compliance of the structure is 8.503. The structure can also be analyzed by using 19×44 regular finite elements, as shown in Fig. 1(c), in which $x_i$ for "grey" elements is 0.5, so that both models have the same volume. The calculated compliance of the structure depends on the used material model: 8.535 for the ersatz material model (error, 0.38%), 9.382 for the HS upper bound model (error, 10.34%) and 9.913 for the SIMP model with $p = 3$ (error, 16.58%). Through the above analysis, we can conclude that the ersatz material model obtains the most accurate solution compared with the HS upper bound model and the SIMP model in the case that a structure contains intermediate boundary elements. Physically, those elements are partly covered by the solid and $x_i$ ($x_{min} < x_i < 1$) should be interpreted as volume fractions of elements, which can be adequately simulated by the ersatz material model [1, 2, 4]. Therefore, the ersatz material model suits for topology optimization aiming at a smooth design defined in eqn. (2). Nevertheless, the HS upper bound material model and the SIMP model suggest that $x_i$ denotes the density of an element filling with an optimal or artificial isotropic porous material. As a result, both the HS upper bound material model and the SIMP model overestimate the compliance of the structure due to boundary 'weak' elements. Ideally, all intermediate elements should be excluded from the final design so that the error of finite element analysis can be avoided. Therefore, a material penalization scheme only suits for topology optimization seeking a 0/1 design without boundary intermediate elements as defined in eqn. (3).

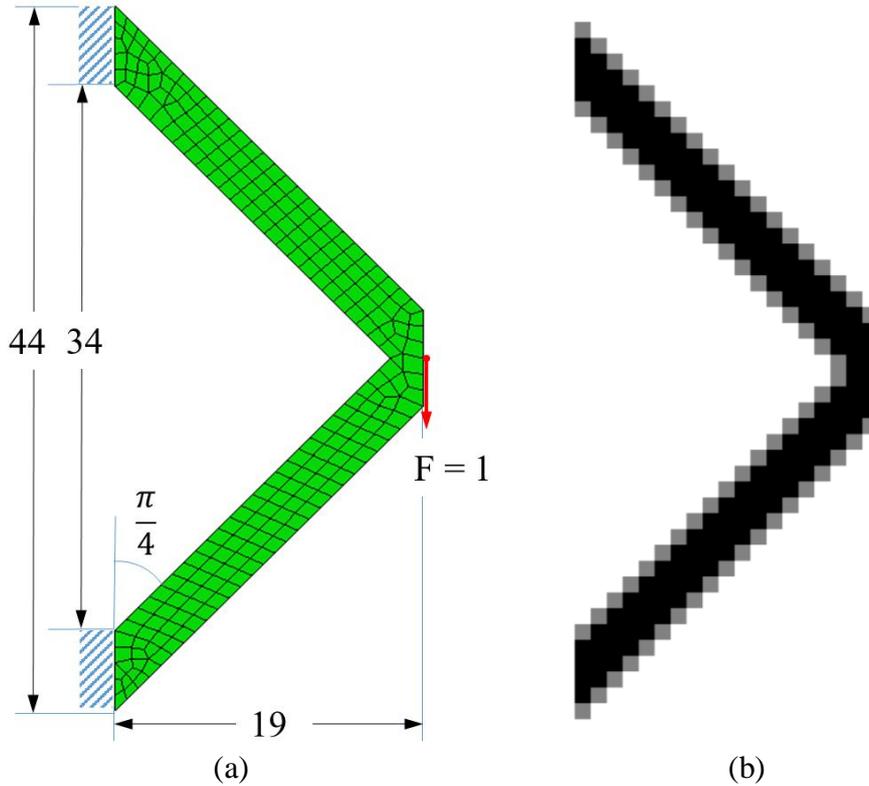

(a) (b)
Fig. 3. A smooth structure simulated with (a) the irregular mesh in ABAQUS; (b) the fixed-mesh.

To illustrate the strategy of the current element-based topology method, we take a spring model with two design variables shown in Fig. 4(a) as an example. The optimization aims to minimize the compliance and the problem is stated by

$$\begin{cases} \min: f(x_1, x_2) = (x_1 + 0.9)^{-1} + (x_2 + 1)^{-1} \\ \text{s.t.:} \begin{cases} x_1 + x_2 = 1 \\ 0 \le x_1, x_2 \le 1 \end{cases} \end{cases} \quad (5)$$

By substituting the constraint, $x_2 = 1 - x_1$, into the objective function, the variation of the objective function against $x_1$ can be plotted with the solid concave curve in Fig. 4. Under this circumstance, the optimal solution, $x_1 = 0.55$, is a non-0/1 solution. In order to achieve a 0/1 design, the stiffness of the springs can be penalized as $k_i = x_i^p$ ($p > 1$) like the SIMP model, the values of the objective function are modified accordingly except for those at $x_1 = 0$ and 1 as shown by the dot lines in Fig. 4 for $p = 1.5$. As a result, the location of the minimum objective function is changed. When $p$ is large enough, such as $p =$ 2 in this example, the minimum objective function automatically corresponds to a 0/1 solution ($x_1 = 1$ and $x_2 = 0$) by considering their upper and lower limits. This example well demonstrates how the material penalization elegantly tackles with a discrete optimization problem in eqn. (3) by *abandoning 0/1 constraints of design variables*. However, due to the complexity of topology optimization with a large number of design variables, topology optimization using the material penalization may also take the risk of falling into a local non-0/1 optimum (e,g., the curve for $p = 1.5$ in Fig. 4).

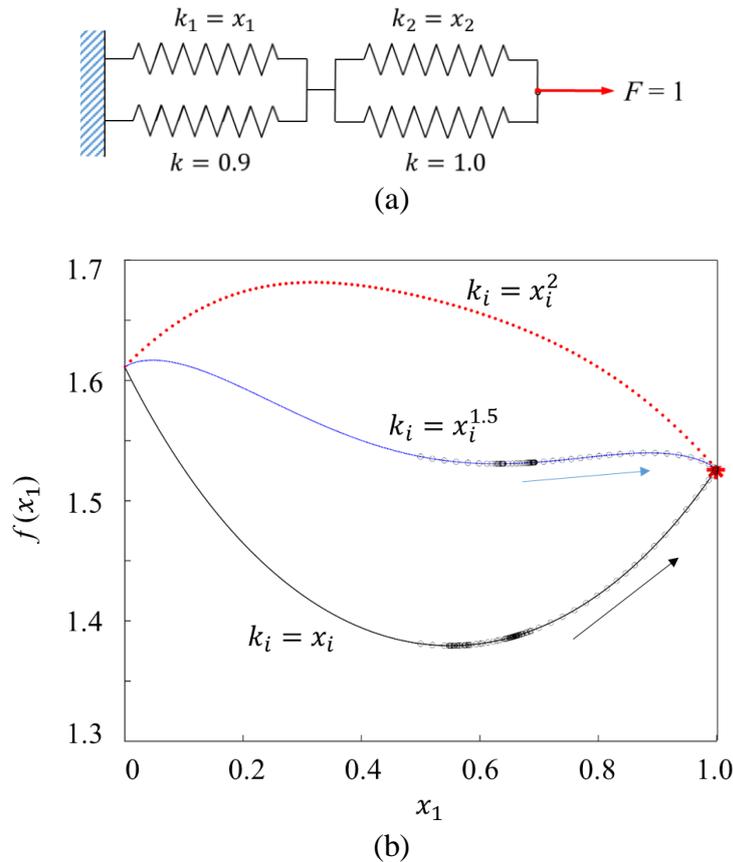

Fig. 4 (a) The spring model for compliance minimization. (b) The variations of the objective function and the modified objective functions using the SIMP model with $p$ =1.5 and 2.0, respectively. The hollow circles denote the search trajectory using the proposed floating projection constraint in Section 3.

It is well known that topology optimization using the ersatz material model becomes a so-called "variable-thickness-sheet" problem [35], which results in a non-0/1 solution, as shown in Fig. 4(b). Both smooth and 0/1 designs of structures are essentially topological solutions (perforated plates in 2D cases) except that the degree of approaching 0/1 is different. In this regard, adopting the 0/1 constraints of design variables seems to be one possible way to achieve a smooth design with intermediate boundary elements, in which the ersatz material model should be adopted to avoid the error of the fixed-mesh finite element analysis due to material penalization. It is also an alternative way to avoid a local non-0/1 optimum and achieve a pure 0/1 design for topology optimization using a material penalization scheme. However, imposing 0/1 constraints directly for thousands of design variables or more in topology optimization is almost impossible, especially for using the ersatz material model aiming for a smooth design. As for this dilemma, this paper adopts a floating projection constraint to simulate 0/1 constraints, which gradually pushes design variables toward 0 or 1 as indicated by the circles in Fig. 4.

## 3. The floating projection topology optimization (FPTO)

With the relaxed design variables, the topology optimization problems in eqns. (2) and (3) can be solved by modifying the objective function as

$$\min: f = C + \Lambda(V - V_0) + \sum \Lambda_j (g_j - g_j^*) \tag{6}$$

Where $\Lambda \geq 0$ and $\Lambda_j \geq 0$ are the Lagrange multipliers of the corresponding constraints. $g_j$ and $g_j^*$ consider the discrete 0/1 constraints of the design variables, which include the explicit upper and lower bound constraints, implicit filter and floating projection constraints. Since those explicit and implicit constraints will be enforced in each iteration, the last term in eqn. (6) can be ignored. Thus, the optimality criterion (OC) can be expressed by

$$\frac{\partial f}{\partial x_i} = \frac{\partial C}{\partial x_i} + \Lambda \frac{\partial V}{\partial x_i} = 0 \tag{7}$$

The sensitivity of the compliance for each element can be easily calculated by [8]

$$\frac{\partial C}{\partial x_i} = -\mathbf{u}^\mathrm{T} \frac{\partial \mathbf{K}}{\partial x_i} \mathbf{u} = \begin{cases} -\mathbf{u}_i^\mathrm{T} \mathbf{K}^1 \mathbf{u}_i & \text{for the ersatz material} \\ -3(3 - 2x_i)^{-2} \mathbf{u}_i^\mathrm{T} \mathbf{K}^1 \mathbf{u}_i & \text{for HS upper bound (in 2D)} \\ -p x_i^{p-1} \mathbf{u}_i^\mathrm{T} \mathbf{K}^1 \mathbf{u}_i & \text{for the SIMP model} \end{cases} \tag{8}$$

where $\mathbf{K}^1 = \int \mathbf{B}^\mathrm{T} \mathbf{D} \mathbf{B} dV$ is the elemental stiffness matrix of the solid. Van Dijk et al. [14] has reviewed the strong similarities between naturally extended velocity fields based on the shape gradient in the level-set method and the sensitivities in the element-based topology optimization. When the shape gradient is equivalent to the sensitivity for the ersatz material model in eqn. (8), the level-set method is only applicable for shape optimization by the boundary evolution [36]. To nucleate holes in the design domain, the concept of the material penalization should be explicitly or implicitly utilized, e.g., the density-based level-set method [37] and the parameterized level-set method [18]. Note that the elemental strain energy used as the update criterion for the design variables is equivalent to the sensitivity for the SIMP model with $p = 2$ [14]. To my best knowledge, topology optimization using the sensitivity for the ersatz material model in eqn. (8) is still unavailable except for the proposed FPTO method in this paper.

To damping the update of the design variables, the elemental sensitivities can be averaged with their values in the previous iterations [10] as

$$\frac{\partial C^{it}}{\partial x_i} = \frac{\frac{\partial C^{(it-1)}}{\partial x_i} + \frac{\partial C}{\partial x_i}}{2} \qquad \text{when } it > 1 \tag{9}$$

where the superscript $it$ denotes the current iteration number. The sensitivity of the volume is

$$\frac{\partial V}{\partial x_i} = V_i \tag{10}$$

Hence, the optimality criterion, eqn. (7) can be rewritten with

$$-\frac{\partial C^{it}/\partial x_i}{\Lambda \, \partial V/\partial x_i} = 1 \tag{11}$$

Thus, the design variables can be updated according to

$$x_i^1 = \left(-\frac{\partial C^{it}/\partial x_i}{\Lambda \, \partial V/\partial x_i}\right) x_i^{it-1} \tag{12}$$

However, the solution satisfying the optimality criterion in eqn. (7) may not be a smooth or 0/1 design, that topology optimization aims at, as illustrated in Fig. 4. Therefore, the design variables will be modified to enforce a series of explicit and implicit constraints. First, the design variables are modified by satisfying the upper and lower bounds

$$x_i^2 = \begin{cases} x_{min} & \text{if } x_i^1 \leq x_{min} \\ x_i^1 & \text{otherwise} \\ 1 & \text{if } x_i^1 \geq 1 \end{cases} \tag{13}$$

Topology optimization of continuum structures often encounters numerical instabilities, e.g. the checkerboard pattern [8, 38]. Here, the following filter applies an implicit constraint on the design variables as

$$x_i^3 = \frac{\sum w(r_{ij}) x_j^2}{\sum w(r_{ij})} \tag{14}$$

where $r_{ij}$ denotes the distance between the centres of the *i*th and *j*th elements. $w(r_{ij})$ is the weight factor defined by

$$w(r_{ij}) = \begin{cases} r_{min} - r_{ij} & \text{if } r_{ij} \leq r_{min} \\ 0 & \text{if } r_{ij} > r_{min} \end{cases}$$

where $r_{min}$ is the filter radius specified by the user. Lastly, in order to push the design variables toward 0 or 1, the following floating projection constraint based on the Heaviside function is further applied to the design variables as

$$x_i^{it} = \frac{\tanh(\beta \cdot th^{it}) + \tanh[\beta \cdot (x_i^3 - th^{it})]}{\tanh(\beta \cdot th^{it}) + \tanh[\beta \cdot (1 - th^{it})]} \tag{16}$$

where $\beta > 0$ controls the steepness of the Heaviside function. The floating threshold, $th^{it}$, can be determined by ensuring that summations of design variables before and after the projection are equal, i.e., $\sum x_i^{it} = \sum x_i^3$. Meanwhile, the Lagrange multiplier, $\Lambda$, can be determined by the volume constraint, $\sum x_i^{it} = V_0$, according to the bi-section method [39] or the Newton-Raphson method [40]. The updated design variables, $\pmb{x}^{it} = \{x_1^{it}, x_2^{it}, \dots, x_i^{it}, \dots, x_N^{it}\}$ will be used for the next finite element analysis. For some cases, one may instead use a moving limit $\delta$, so that $x_{min} \leq x_i^{it-1}(1-\delta) \leq x_i^{it} \leq x_i^{it-1}(1+\delta) \leq 1$. $\delta = 0.02$ is used for all numerical examples in this paper.

To gradually push the design variables toward 0 or 1, $\beta$ may start from a small positive value, e.g. $10^{-6}$ and then increases with $\Delta\beta$ once the following convergence criterion is satisfied

$$\epsilon = \frac{\sum_{i=1}^{N} |x_i^{it} - x_i^{it-1}|}{N} \leq 10^{-3} \tag{17}$$

The final steepness of the floating projection, $\beta$, controlls the degree of approaching a 0/1 solution in the FPTO method. $\beta$ will be stopped to increase once the solution is close to a smooth design, which will be explained in the next section. Then, the whole optimization is stopped until the strict convergence criterion, $\epsilon \leq 10^{-4}$, is satisfied.

When the HS upper bound model or the SIMP model is used, the FPTO method aims for a pure 0/1 design. Thus, we will increase $\beta$ once $\epsilon \leq 10^{-3}$. The whole optimization will be stopped when $\epsilon \leq 10^{-5}$, which ensures that $\beta$ is large enough for achieving a nearly 0/1 design.

## 4. Representation of smooth boundaries

To represent an element-based design with smooth boundaries, the design variables of elements $\boldsymbol{x}^{it}$ can be linearly interpolated into the whole design domain, $\boldsymbol{x}^{it}(x, y)$. $\boldsymbol{x}^{it}(x, y)$ can be assumed to be a pixel-based design, such as 20×20 pixels within each element in this paper. Then, a level-set function $\phi^{it}(x, y) = \boldsymbol{x}^{it}(x, y) - ls^{it}$ can be constructed, where $ls^{it}$ is the threshold ensuring that the volume of the 0/1 pixel-based design is equal to that of the element-based design. The smooth boundary is represented by the zero level-set as

$$\begin{cases} \phi^{it}(x, y) > 0 & \text{for solid region} \\ \phi^{it}(x, y) = 0 & \text{for boundary} \\ \phi^{it}(x, y) < 0 & \text{for void region} \end{cases} \quad (18)$$

Next, the 0/1 pixel-based design projects back to the fixed mesh to calculate volume fractions of all elements, $\boldsymbol{v}^{it} = \{v_1^{it}, v_2^{it}, \dots, v_i^{it}, \dots, v_N^{it}\}$. The smooth design under the fixed mesh can be accurately expressed by $\boldsymbol{v}^{it}$ with one layer of intermediate boundary elements, whose material properties should be interpolated by the ersatz material model, as explained in Section 2.

When the solution for a given $\beta$ is convergent, we need to check if the optimized element-based design, $\boldsymbol{x}^{it}$, approaches a smooth design, $\boldsymbol{v}^{it}$. One way is to comparing the geometrical difference between the optimized element-based design with the smooth design [41]. In this paper, one additional finite element analysis is conducted to calculate the compliance of the smooth design, $C(\boldsymbol{v}^{it})$. The difference of the objective function between an optimized element-based design and the corresponding smooth design can be measured by

$$\tau = \left|\frac{C(\boldsymbol{v}^{it}) - C(\boldsymbol{x}^{it})}{C(\boldsymbol{x}^{it})}\right| \qquad (19)$$

When $\tau \leq 10^{-2}$, it indicates that an optimized element-based design is almost equivalent to a smooth design. In other words, the desired 0/1 level of an optimized element-based design is achieved, and $\beta$ will no longer increase.

## 5. Numerical Examples

In this section, we will present some numerical examples to demonstrate the effectiveness of the FPTO method for achieving smooth or 0/1 designs. In all examples, the material properties of the solid are Young's modulus $E^1 = 1$ and Poisson's ratio $v = 0.3$, and the filter radius is $r_{min} = 2$. $\Delta\beta = 1$ is used for all examples.

5.1 Smooth designs based on the ersatz material model

5.1.1 Example 1

The first example is on the design of a 2D cantilever with 120 in length and 80 in height. A unit force is applied at the center of the right edge. The whole design domain is divided into 120×80 four-node plane-stress elements. The final volume is restricted to be no more than 50% of the whole design domain.

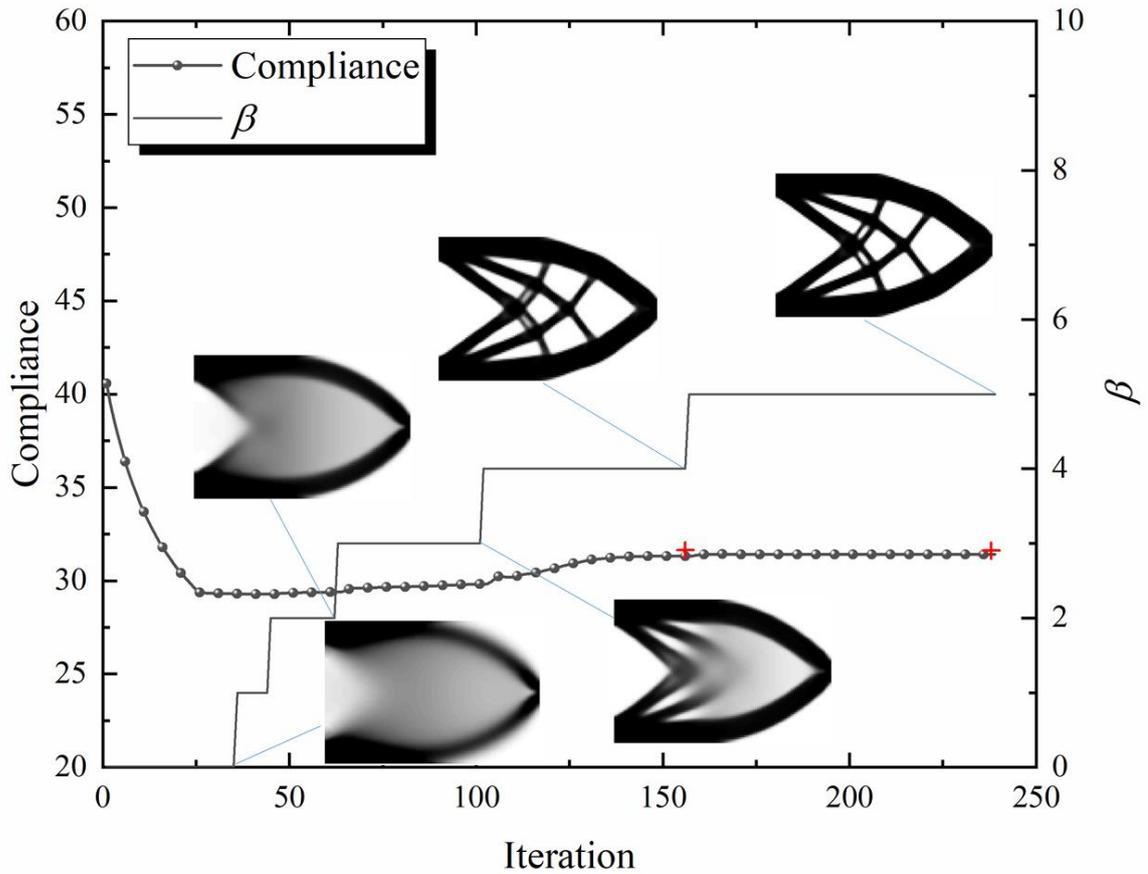

Fig. 5 Evolution histories of compliance, topology and $\beta$ for the FPTO method using the ersatz material model (red marks "+" denote the compliances of smooth designs).

When the proposed FPTO method is applied, the evolution histories of compliance, topology, and $\beta$ are shown in Fig. 5. When $\beta = 10^{-6}$, the effect of the floating projection constraint can be ignored. The design converges at the well-known 'variable-thickness-sheet' solution [35] with a large volume of 'grey' elements as shown in the inset of Fig. 5 and its compliance is 29.29. In order to escape from this non-0/1 solution and approach a smooth design, the floating projection constraint is activated by gradually increasing $\beta$. Meanwhile, the compliance increases due to the imposed floating projection constraint. When $\beta = 3$, the topology starts to appear but the solution still contains many "grey" elements. When $\beta = 4$, a clear topology is obtained, but the solution does not satisfy $\tau \leq 10^{-2}$ in eqn. (19). It means that the desired 0/1 level for the representation of smooth boundaries has yet achieved, and $\beta$ should further increase to push the design variables toward 0/1. When $\beta$ increases to 5, $\tau \leq 10^{-2}$ is satisfied and the solution stably converges at 31.50.

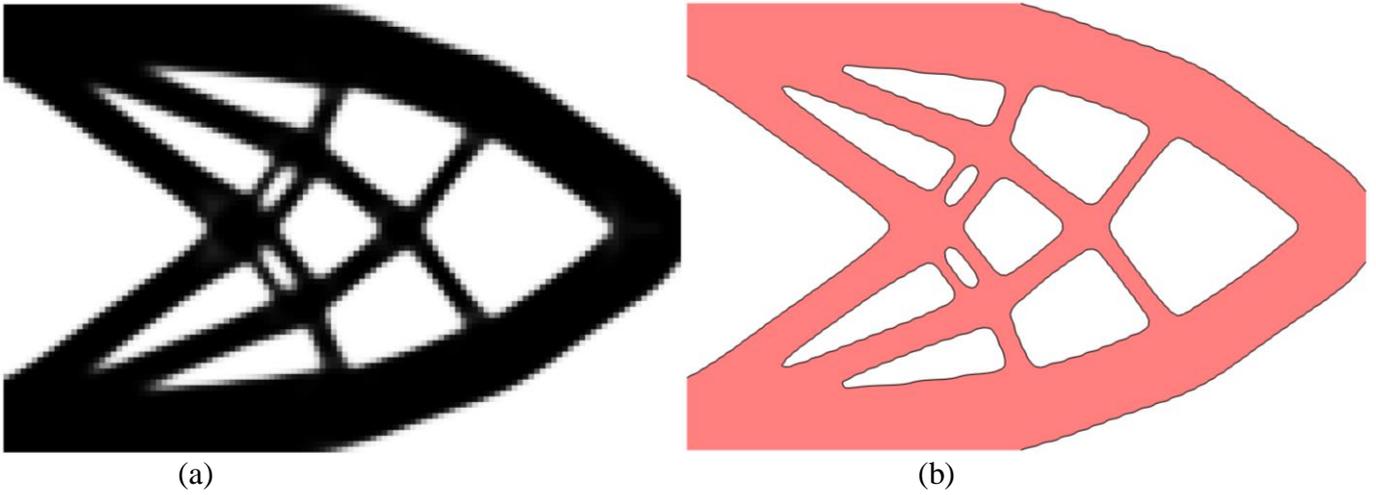

(a)                                 (b)

Fig. 6 (a) Element-based design with $C = 31.41$; (c) The smooth design with $C = 31.61$.

The final element-based design is shown in Fig. 6(a) and the corresponding smooth design is shown in Fig. 6(b). The compliance of the smooth design is $C = 31.61$, which is close to that of the element-based design, 31.41. The relative difference of the objective function is $\tau = 0.64\%$.

For the convenience of the reader, a Matlab code for this example is attached in the Appendix. The code can easily be modified for other 2D examples.

5.1.2 Example 2

The second example is on the design of an MBB beam with 240 in length and 40 in height. A unit force is applied in the middle of the top edge. The whole design domain is meshed with 240×40 four-node plane-stress elements. The volume is constrained to being no more than 50% of the whole design domain.

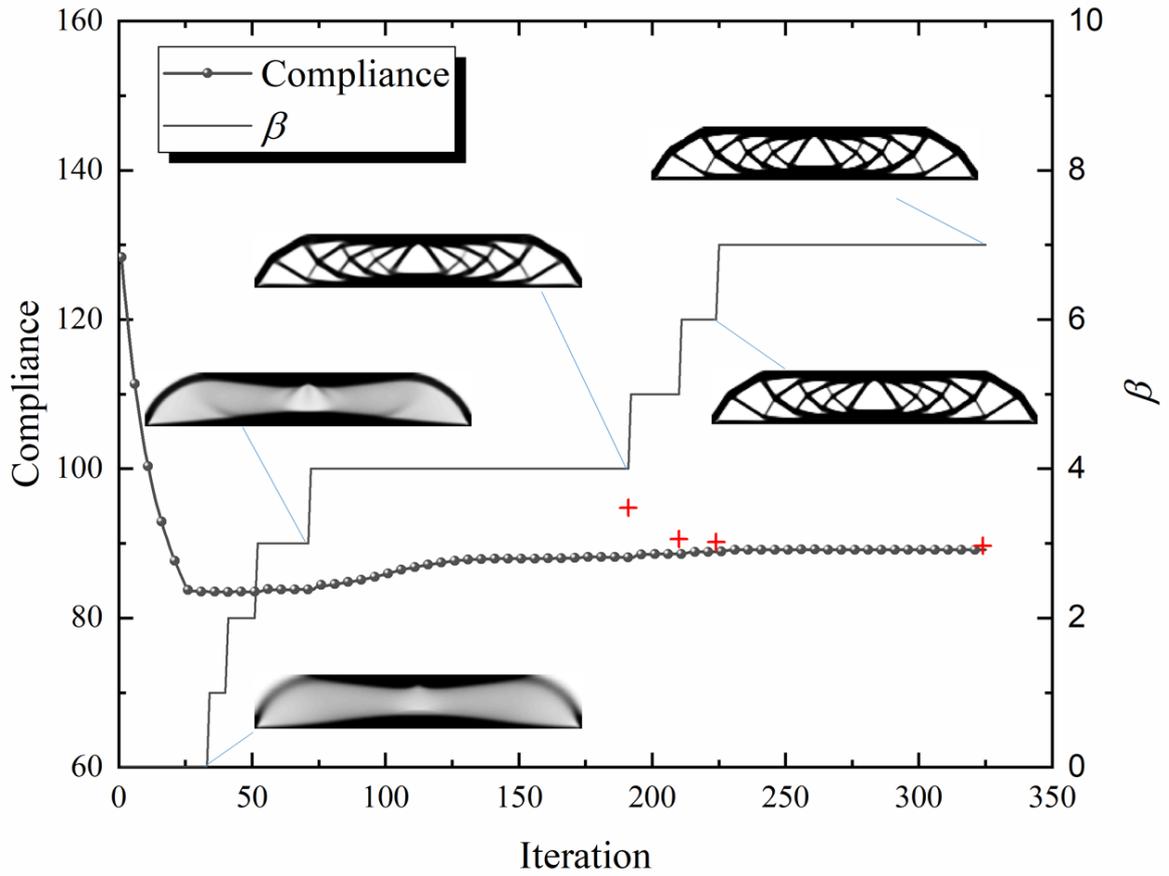

Fig. 7 Evolution histories of compliance, topology and $\beta$ for the FPTO method using the ersatz material model (red marks "+" denote the compliances of smooth designs).

Fig. 7 shows the evolution histories of compliance, topology and $\beta$ for the FPTO method using the ersatz material model for the MBB beam. The compliance initially decreases as $\beta = 10^{-6}$, and reaches its minimum value. The corresponding design contains a large volume of "grey" elements. Then the compliance starts to increase as $\beta$ increases. An adequate topology is formed as $\beta=4$, and meanwhile the compliance of the smooth design approaches to that of the element-based design. When $\beta$ increases to 7, the element-based design achieves the desired 0/1 level, and the compliance of the element-based design converges at 89.46. Fig. 8 shows the comparison of the final element-based design and its smooth design, where the measured difference, $\tau = 0.25\%$, is marginal. The enlarged member of the element-based design, as the inset of Fig. 8(a), indicates that two boundaries nearly meet together. Even so, the constructed level-set function successfully identifies a thin member, as shown in Fig. 8(b). The width of the member is even smaller than the size of an element. It is impossible to forming such a thin member in a 0/1 design.

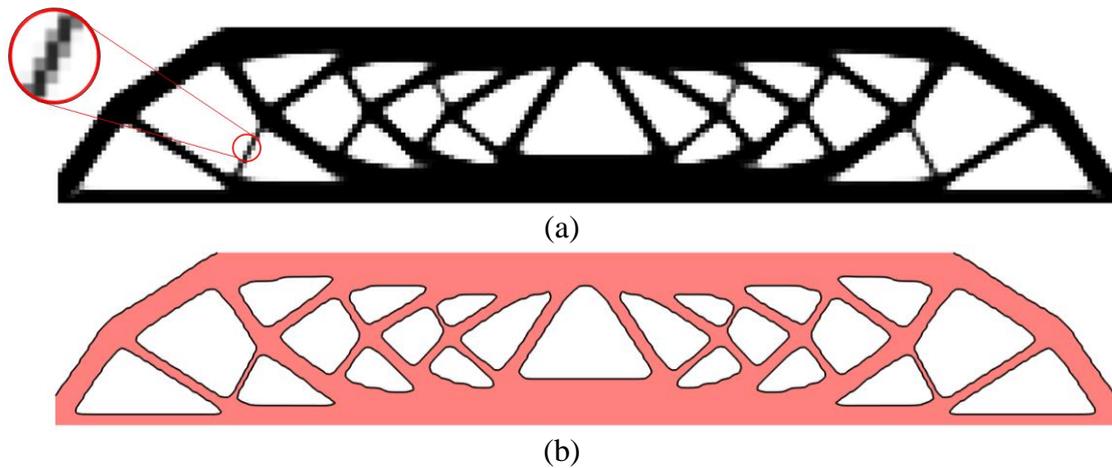

(a)

(b)

Fig. 8 (a) Optimized element-based design, $C$ = 89.46; (c) The smooth design, $C$ = 89.68.

The above two examples demonstrate that the structural topologies result from the imposed floating projection constraint. This formation mechanism of structural topologies is fundamentally different from those through material penalization or boundary evolution and provides an alternative way to solve topology optimization problems. Compared with the LS method [2-5], the FPTO algorithm can be easily implemented in the framework of the fixed-mesh finite element analysis, and the resulting smooth design is less dependent on an initial guess. Nevertheless, we also noticed that the formation of structural topology is relatively slow, and therefore topology optimization using the ersatz material model is not recommended for seeking a 0/1 design.

5.2 0/1 designs based on material penalization

As discussed in Section 2, topology optimization using the material penalization scheme is more suitable for a 0/1 design. Next, we will revisit the above examples using the FPTO method based on the SIMP model ($p$ = 3) and the HS upper bound model. In the final design, elements with $x_{min} + 0.01 \leq x_i \leq 0.99$ are assumed as intermediate elements.

5.2.1 Example 1

The FPTO method based on the SIMP model with $p=3$ is applied for the design of the 120×80 cantilever. The evolution histories of compliance, topology and $\beta$ are shown in Fig. 9. It can be seen that the topology is identified for $\beta = 10^{-6}$, which is different from that using the ersatz material model. This indicates that the formation of the topology attributes to the material penalization, rather than the floating projection constraint. This design belongs to a local non-0/1 optimum evidenced by the further decrease of the compliance as increasing $\beta$ in the floating projection and pushing intermediate elements towards 0 or 1. For each $\beta$, the solution converges in a few iterations. Since $\beta$ does not affect the finite element analysis, it can be increased to an extremely large value till achieving a pure 0/1 solution. Finally, the compliance of the optimized cantilever converges at a minimum value, 32.39.

It is noticed that the above optimization also experiences a design with one layer of intermediate boundary elements. For example, the convergent design for $\beta = 10^{-6}$ can also be processed for a smooth design by the proposed level-set function. However, the compliance of the corresponding smooth design is 32.08, as marked in Fig. 9, which greatly deviates from the evaluated value of the element-based design, 34.56 ($\tau = 7.7\%$). This discrepancy is mainly because the smooth design is analysed using the ersatz material model. One may argue that the difference, $\tau < 1\%$, can also be obtained if the smooth design is analysed using the same SIMP model. Theoretically, this is unacceptable because the evaluated compliance is much higher than the actual one of the smooth design, as explained in Section 2. This discrepancy also indicates that the displacement field in optimization may be unreliable even that a design contains one layer of intermediate boundary elements. More seriously, provided that any constraints are evaluated by the displacement field, a smooth design produced in this way may violate the actual constraints. Therefore, topology optimization based on material penalization is recommended for seeking a 0/1 design rather than a smooth design.
.

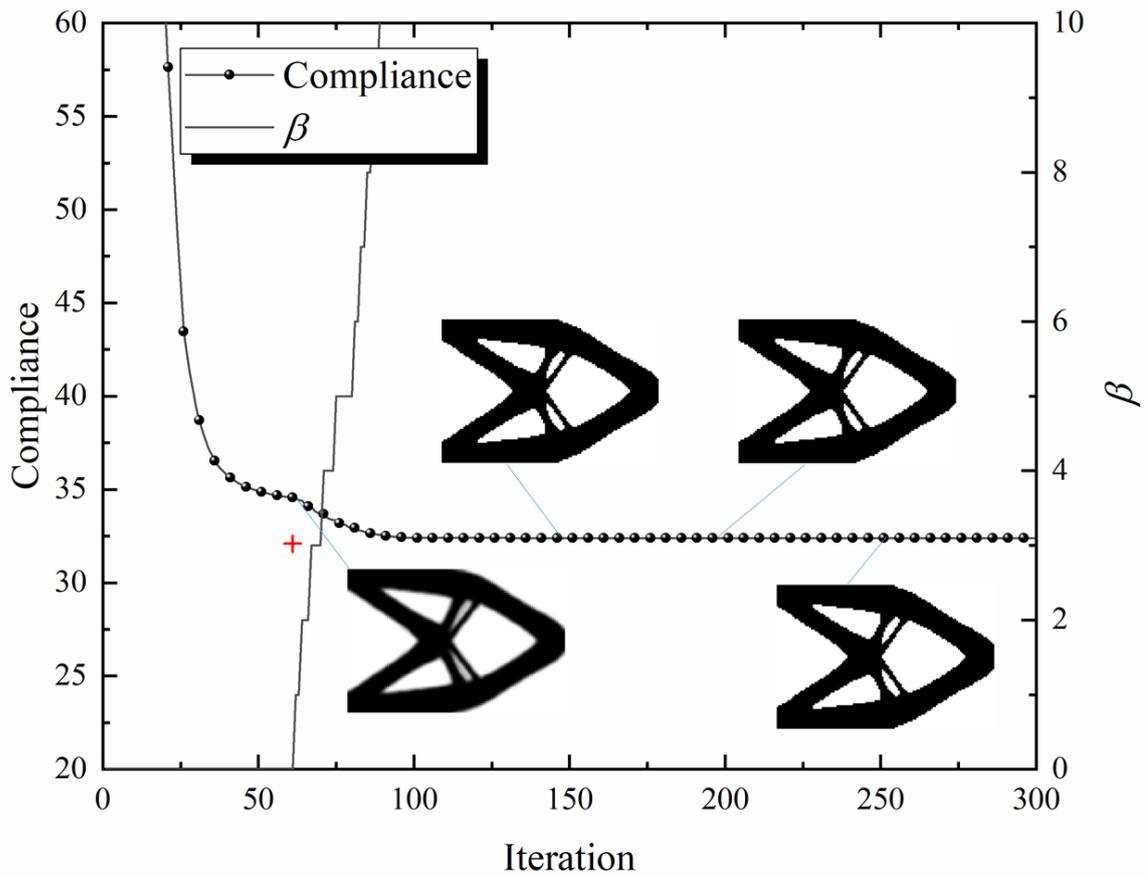

Fig. 9 Evolution histories of compliance, topology, and $\beta$ for the FPTO method using the SIMP model with $p = 3$ (the red mark "+" denotes the compliance of the corresponding smooth design).

Fig. 10 shows the final design resulting from the FPTO method using the SIMP model with $p = 3$ and the HS upper bound model, respectively. Both designs are almost pure 0/1 solutions, e.g., 12 intermediate elements in Fig. 10(a) and 6 in Fig. 10(b). Although both designs have different topologies, their compliances are very close. The compliance of those 0/1 designs is higher than that of the FPTO design using the ersatz material model.

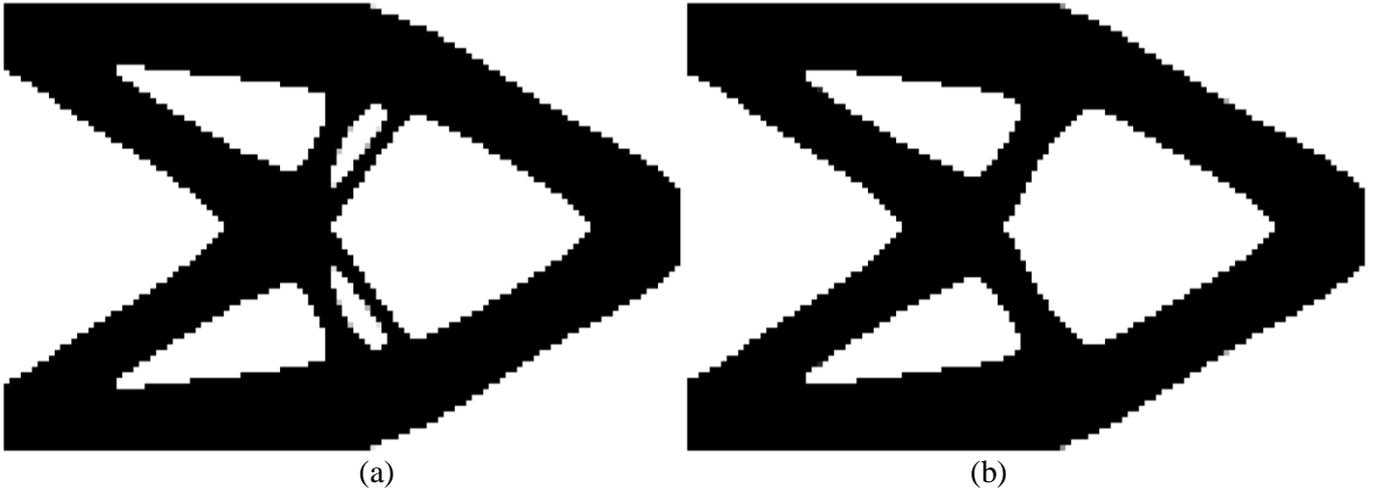

(a)                                                              (b)

Fig. 10 (a) FPTO 0/1 design using the SIMP model with $p = 3$ (12 intermediate elements) and $C = 32.39$; (b) FPTO 0/1 design using the HS upper bound model (6 intermediate elements) and $C = 32.41$.

5.2.2 Example 2

Next, we employ the FPTO algorithm for designing the 240×40 MBB beam using the SIMP model with $p = 3$ for a 0/1 solution. Fig. 11 shows the evolution histories of compliance, topology, and $\beta$. The variation trends of compliance and $\beta$ are very similar to those in the above example. Once again, an adequate topology is formed before the active of the floating projection constraint. It further confirms that the formation of topology mainly attributes to material penalization, and the floating projection constraint overrules local non-0/1 optima and pushes intermediate elements towards 0 or 1. The compliance of the element-based design for $\beta = 10^{-6}$ is 103.83 which is much larger than that of the corresponding smooth design, 91.24 ($\tau = 13.8\%$). This further confirms that topology optimization based on material penalization is not applicable for seeking a smooth design.

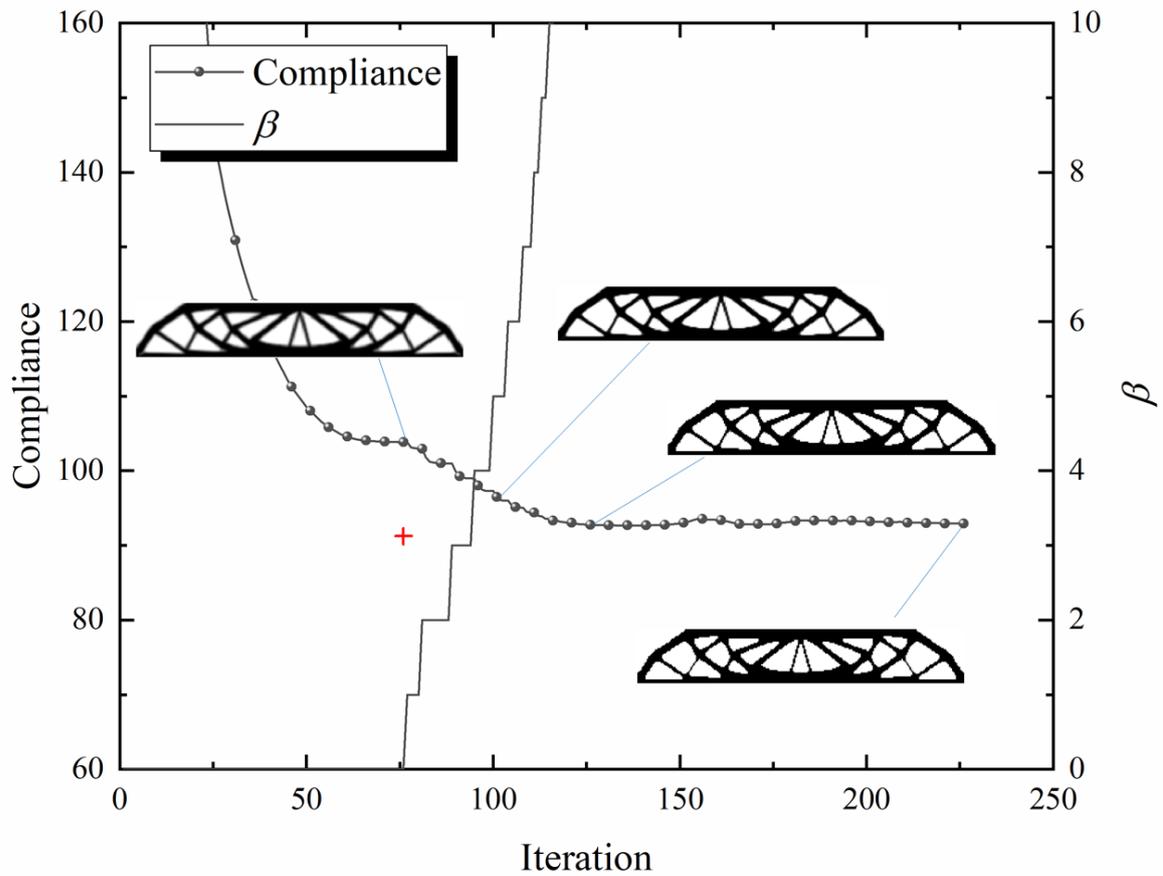

Fig. 11 Evolution histories of compliance, topology and $\beta$ for the FPTO method using the SIMP model with $p = 3$ (the red mark "+" denotes the compliance of the corresponding smooth design).

The final designs resulting from the FPTO method based on the SIMP model with $p = 3$ and the HS upper bound model are shown in Figs. 12(a) and (b), respectively. Both designs are close to pure 0/1 solutions. The 16 intermediate elements in Fig. 12(a) mainly distribute at the two narrowest members to avoid the breakage of those members. Although the topologies of the two designs are different, their compliances are close. Those values are also higher than that of the FPTO design using the ersatz material model.

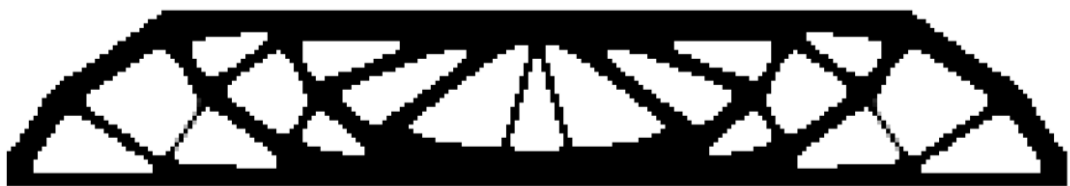

(a)

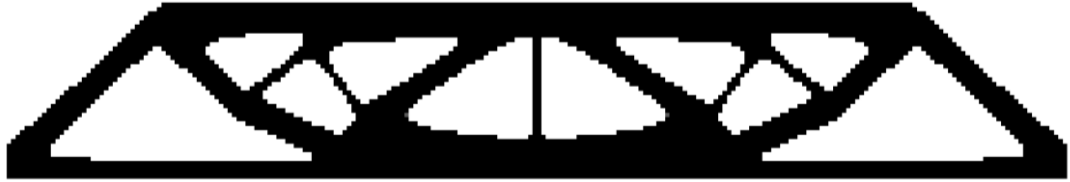

(b)

Fig. 12 FPTO 0/1 design using the SIMP model with $p = 3$ (16 intermediate elements) and $C = 92.91$; (b) FPTO 0/1 design using the HS upper bound model (2 intermediate elements) and $C = 92.77$.

5.3 Verification and Discussion

In order to verify those smooth and 0/1 designs, the above topology optimization problems are solved by the BESO method based on the SIMP model with $p = 3$ [30, 39] and the continuation method [17]. BESO starts from full design and gradually decreases the volume fraction with the evolution rate, 0.02. For the continuation method, $p = 1$ is assigned initially and then gradually increases with $\Delta p = 0.2$. Other parameters are the same as those used in the FPTO method. It should note that both BESO and the continuation method aim for seeking a 0/1 design.

Figs. 13 and 14 show the optimized designs and their compliances resulting from the BESO method and the continuation method, for the cantilever and the MBB beam, respectively. The compliances of the BESO designs are very close to those of the FPTO 0/1 designs, shown in Fig. 10 for the cantilever and Fig. 12 for the MBB beam. However, those BESO and FPTO designs have different topologies and their compliances are always higher than those of the optimized designs resulting from the continuation method.

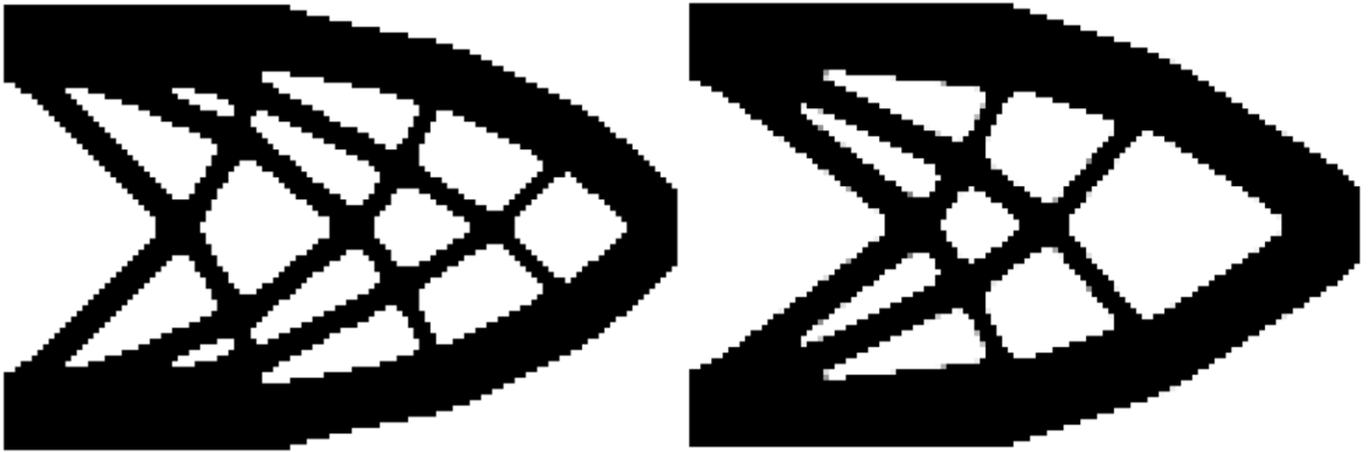

(b)                                                    (b)

Fig. 13 (a) The BESO design based on the SIMP model with $p = 3$, C = 32.49; (b) The design resulting from the continuation method, $C = 32.04$.

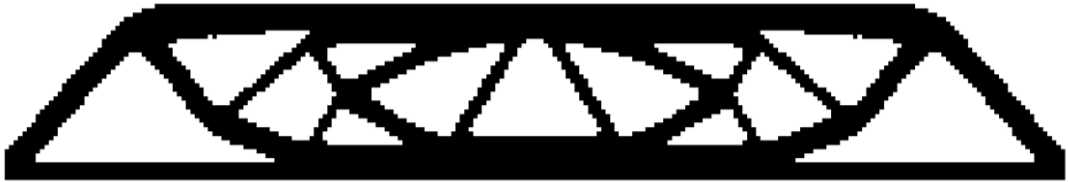

(a)

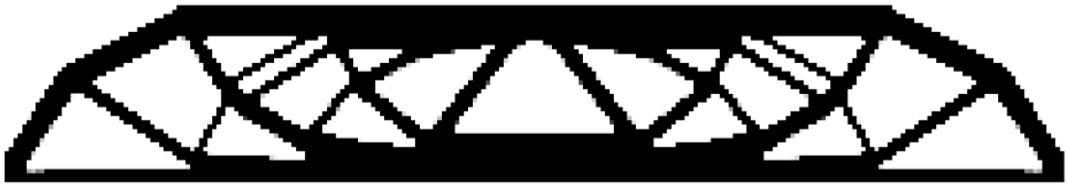

(b)

Fig. 14 (a) The BESO design based on the SIMP model with $p = 3$, $C = 92.80$; (b) The design resulting from the continuation method, $C = 91.07$.

Theoretically, the continuation method is able to eliminate the effect of the material interpolation scheme and achieve a global 0/1 optimum. It is not surprising that its compliance is always lower than those of other 0/1 designs for a given material penalization scheme. Next, we compare the optimized designs resulting from the continuation method with the FPTO smooth designs using the ersatz material model in Sections 5.1. For the design of the cantilever, the optimized topology of the FPTO smooth design using the ersatz material model in Fig. 6(b) is essentially the same as that of the continuation method, except that one member is split into two thin members in the FPTO design. For the MBB beam, both designs are fundamentally similar except that the optimized topology from the continuation method shows fewer members. As explained in Section 5.1.2, the FPTO design using the ersatz material model enables to capture of thin members less than the size of one element, which is impossible for the continuation method aiming at a 0/1 design. As a result, the optimized design from the continuation method has fewer members than that of the FPTO smooth design. As for the objective function, the compliance of the FPTO smooth design is even lower than that of the 0/1 design from the continuation method. The effectiveness and capabiblity of the FPTO method for other topology optimization problems, such as frequency optimization and the design of compliant mechanisms, can refer to the reference [41].

The above comparsion well demonstrates the effectiveness of the proposed FPTO method for achieving a smooth or 0/1 design. Since the smooth design has excellent aesthetic characteristics that can be exported to CAD for fabrication, the FPTO method using the ersatz material model is highly recommended, especially for 3D structures. For example, the design domain, boundary and loading conditions for a 3D structure are given in Fig. 15(a). The volume of the final design is restricted to be less than 15% that of the design domain. The evolution histories of the objective function and $\beta$ are given in Fig. 15(b), where the compliances of smooth designs for each $\beta$ are also marked with symbols '+'. It can be seen that the compliance of the smooth design for $\beta = 4$ approachs that of the element-based design. This indicates that the structural topology has been successfully formed. The whole optimization is stopped after 164 iterations. The final element-based design shown in Fig. 15(c) only displays elements whose design variables are larger than the threshold as illustrated in Section 4. The final smooth design is shown in Fig. 15(d) and its compliance is 4.277, which is only 0.77% above that of the element-based design. Compared with the element-based design in Fig. 15(c), the smooth design in Fig. 15(d) is more attractive for fabrication.

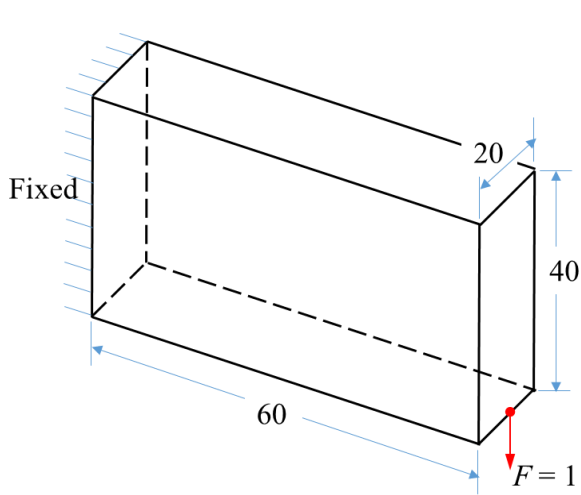 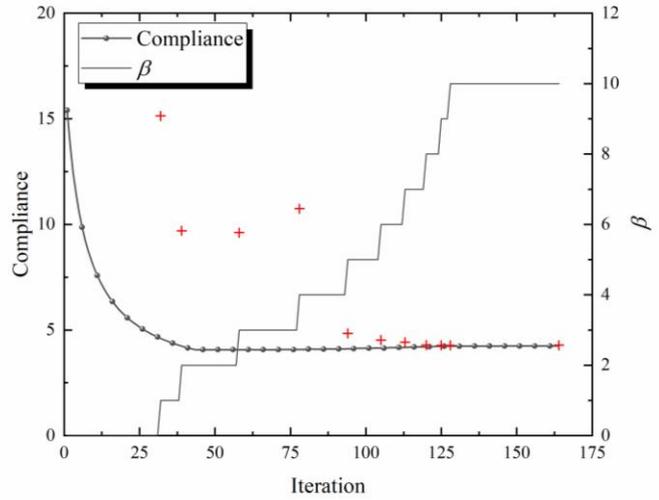

(a)                      (b)

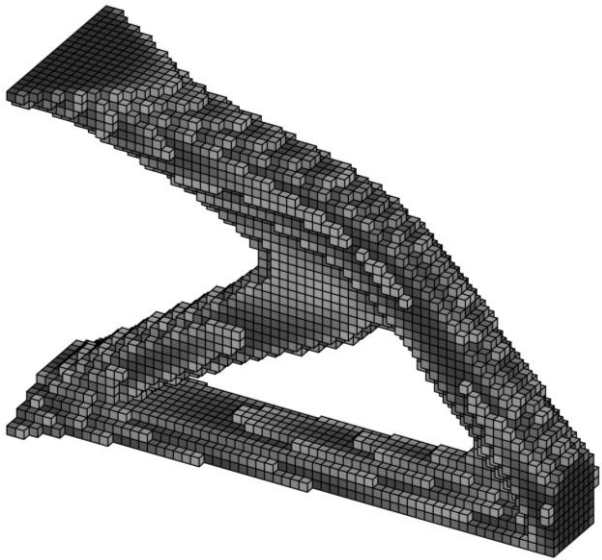 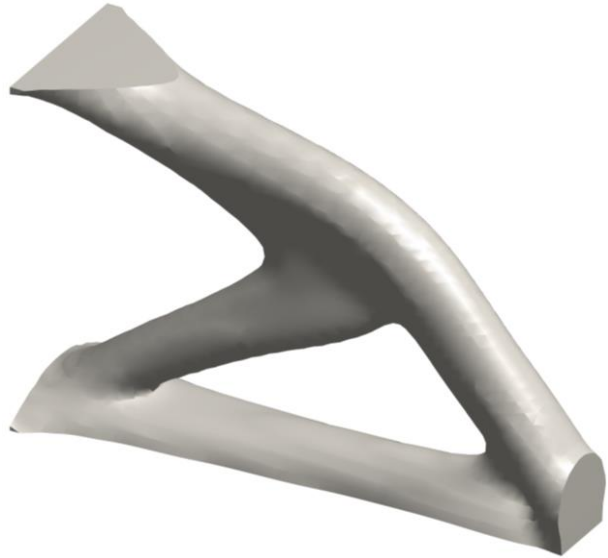

(c)                      (d)

Fig. 15 (a) design domain of a 3D structure; (b) evolution histories of the objective function and $\beta$ (red marks "+" denote the compliance of smooth designs); (c) element-based design with $C = 4.244$; (d) smooth design with $C = 4.277$.

## 6 Conclusions

In this paper, we conduct numerical experiments for a smooth design under the fixed-mesh finite element analysis and conclude that the ersatz material model should be adopt for topology optimization seeking a smooth design and a material penalization model for a 0/1 design. We further demonstrate the capability of the FPTO method by employing various material interpolation schemes, where the proposed floating projection constraint gradually pushes design variables to the desired 0/1 level for a smooth or 0/1 design. The FPTO algorithm is easy to be implemented in the framework of the fixed-mesh finite element analysis. Numerical examples show that the FPTO method based on material penalization is able to overrule any local non-0/1 optima and achieve a nearly pure 0/1 design. The FPTO method based on the ersatz material model produces an equivalent smooth design with one layer of boundary intermediate elements, whose compliance is meanwhile evaluated during optimization. In the case, the formation of structural topology contributing from the floating projection constraint is fundamentally different from the existing topology optimization approaches based on material interpolation or boundary evolution. Compared with a 0/1 design, a smooth design has not only a better structural performance but also has attractive aesthetic characteristics, which can be directly exported to CAD for fabrication.

## Acknowledgments

The author wishes to acknowledge the financial support from Australian Research Council under the Future Fellowship scheme (FT130101094).

# Appendix

```matlab
%% FLOATING PROJECTION TOPOLOGY OPTIMIZATION (FPTO) %%
%% The code seeks a smooth design using the erastz material model
function FPTO2D(nelx,nely,volfrac,rmin)
%% INITIALIZATION
rho = ones(nely,nelx)*volfrac; xmin = 1.e-3;
beta = 1.e-6; inc = 1;  move = 0.02; errmax = 1.e-2;
%% MATERIAL PROPERTIES
E0 = 1; nu = 0.3;
%% MESH & POINT GRIDS
ngrid = 20;
[elex,eley] = meshgrid(1:nelx,1:nely);
[fnx,fny] = meshgrid(1:1/ngrid:nelx,1:1/ngrid:nely);
%% PREPARE FINITE ELEMENT ANALYSIS
A11 = [12 3 -6 -3; 3 12 3 0; -6 3 12 -3; -3 0 -3 12];
A12 = [-6 -3 0 3; -3 -6 -3 -6; 0 -3 -6 3; 3 -6 3 -6];
B11 = [-4 3 -2 9; 3 -4 -9 4; -2 -9 -4 -3; 9 4 -3 -4];
B12 = [ 2 -3 4 -9; -3 2 9 -2; 4 9 2 3; -9 -2 3 2];
KE = 1/(1-nu^2)/24*([A11 A12;A12' A11]+nu*[B11 B12;B12' B11]);
nodenrs = reshape(1:(1+nelx)*(1+nely),1+nely,1+ nelx);
edofVec = reshape(2*nodenrs(1:end-1,1:end-1)+1,nelx*nely,1);
edofMat = repmat(edofVec,1,8)+repmat([0 1 2*nely+[2 3 0 1] -2 -1],nelx*nely,1);
iK = reshape(kron(edofMat,ones(8,1))',64*nelx*nely,1);
jK = reshape(kron(edofMat,ones(1,8))',64*nelx*nely,1);
%% DEFINE LOADS AND SUPPORTS FOR A CANTILEVER
F = sparse(2*(nely+1)*(nelx+1)-nely,1,-1,2*(nely+1)*(nelx+1),1);
fixeddofs = (1:2*(nely+1));
U = zeros(2*(nely+1)*(nelx+1),1);
alldofs = (1:2*(nely+1)*(nelx+1));
freedofs = setdiff(alldofs,fixeddofs);
%% PREPARE FILTER
iH = ones(nelx*nely*(2*(ceil(rmin)+1))^2,1);
jH = ones(size(iH)); sH = zeros(size(iH)); sH1 = zeros(size(iH)); k =0;
for i1 = 1:nelx
 for j1 = 1:nely
  e1 = (i1-1)*nely+j1;
  for i2 = max(i1-ceil(rmin),1):min(i1+ceil(rmin),nelx)
   for j2 = max(j1-ceil(rmin),1):min(j1+ceil(rmin),nely)
    e2 = (i2-1)*nely+j2; k = k+1;
    iH(k) = e1; jH(k) = e2;
    sH(k) = max(0,rmin-sqrt((i1-i2)^2+(j1-j2)^2));
   end
  end
 end
end
H = sparse(iH,jH,sH); Hs = sum(H,2);
%% START ITERATION
change = 1; iter = 0; err = 1;
while (change > 1.e-4 || err > errmax)
 iter = iter + 1; oldrho = rho;
 if iter > 1; olddc = dc; end
 %% FE-ANALYSIS
 sK = reshape(KE(:)*(rho(:)'*E0),64*nelx*nely,1);
 K = sparse(iK,jK,sK); K = (K+K')/2;
 U(freedofs) = K(freedofs,freedofs)\F(freedofs);
 %% OBJECTIVE FUNCTION AND SENSITIVITY ANALYSIS
 ce = reshape(sum((U(edofMat)*KE).*U(edofMat),2),nely,nelx);
 c(iter) = sum(sum(rho.*E0.*ce));
 dc =-E0.*ce;
 %% CONSTRAINT FUNCTION AND SENSITIVITY ANALYSIS
 dfdx = ones(1,nelx*nely)./(nelx*nely);
 %% DAMPING
```

```matlab
  if iter > 1; dc = (dc + olddc)/2.; end
  %% INNER LOOP FOR OPTIMIZATION
  lower = 0; upper = 1.e6;
  while upper-lower > 1.0e-6
   lamada = (lower+upper)/2.;
   rho(:) = max(xmin,min(1,oldrho(:).*(lamada.*(-dc(:)./dfdx(:)))));
   rho1 = reshape((H*rho(:)./Hs),nely,nelx);
   %% FLOATING PROJECTION (FP)
   l1 =0; l2 = 1;
   while (l2-l1) > 1.0e-10
    ls = (l1+l2)/2.0;
    rho2 = max(xmin,(tanh(beta*ls)+tanh(beta*(rho1-ls)))/(tanh(beta*ls)...
       +tanh(beta*(1.-ls))));
    if sum(sum(rho2))-sum(sum(rho1)) > 0; l1 = ls; else; l2 = ls; end
   end
   rho(:)=max(xmin,max(oldrho(:)-move,min(1.,min(oldrho(:)+move,rho2(:)))));
   if sum(rho(:))/(nelx*nely) < volfrac; lower=lamada;
   else; upper=lamada; end
  end
  %% CHECK CONVERGENCE
  change = sum(abs(oldrho(:)-rho(:)))/(nely*nelx);
  %% REPRESENTATION WITH SMOOTH BOUNDARY
  rhofine = interp2(elex,eley,oldrho,fnx,fny,'linear');
  lower = 0; upper = 1; cs(iter) = 0;
  while upper-lower > 1.0e-6
   th = (upper+lower)/2; vx = rho;
   for i = 2:nelx-1
    for j = 2:nely-1
     vx(j,i) = 0;
     for i1 = ngrid*(i-1)-ngrid/2+2:ngrid*(i-1)+ngrid/2
      for j1 = ngrid*(j-1)-ngrid/2+2:ngrid*(j-1)+ngrid/2
       vx(j,i) = vx(j,i)+max(sign(rhofine(j1,i1)-th),xmin);
      end
     end
     vx(j,i) = vx(j,i)/(ngrid-1)^2;
    end
   end
   if sum(vx(:))<sum(oldrho(:)); upper = th; else; lower = th; end
  end
  if change < 1.e-3
   sK = reshape(KE(:)*(vx(:)'*E0),64*nelx*nely,1);
   K = sparse(iK,jK,sK); K = (K+K')/2;
   U(freedofs) = K(freedofs,freedofs)\F(freedofs);
   ce = reshape(sum((U(edofMat)*KE).*U(edofMat),2),nely,nelx);
   cs(iter) = sum(sum(vx.*E0.*ce));
   err = abs(cs(iter) - c(iter))/c(iter);
   if err > errmax; beta = beta + inc; end
  end
  fprintf('It.:%3i Obj.:%8.4f Vol.:%4.3f ch.:%4.5f err.:%4.5f beta.:%4.5f\n',...
  iter,c(iter),mean(rho(:)),change,err,beta);
  %% PLOT RESULTS
  subplot(1,3,1),imagesc(-rho,[-1,-xmin]); colormap(gca,'gray');
  axis equal;  axis tight; axis off;
  top = rho-th; subplot(1,3,2), contourf(elex, flipud(eley), top, [0 0] );
  colormap(gca,'winter'); axis equal;  axis tight; axis off;
  subplot(1,3,3),plot(c,'.-');
  if change<1.e-3; hold on; plot(iter,cs(iter),'+'); end
  set (gca,'xgrid','on','ygrid','on'); hold on;
  xlabel('Iteration'); ylabel('C');
  truesize([210,280]); pause(1e-6);
end
```